\documentclass[a4paper]{PoS}

\usepackage{amsmath,amssymb,color,amscd,subfigure}
\usepackage[english]{babel}
\usepackage{graphicx}

\usepackage{slashed}

\newcommand{\ev}[1]{\left < #1 \right >}

\newcommand{\be}{\begin{equation}}
\newcommand{\ee}{\end{equation}}
\newcommand{\bea}{\begin{eqnarray}} 
\newcommand{\eea}{\end{eqnarray}}

\newcommand{\bmp}{\noindent\begin{minipage}{16cm}}

\definecolor{title}{rgb}{0.6,0.067,0.013}
\definecolor{emph}{rgb}{0.18,0.18,0.60}
\definecolor{emph1}{rgb}{0.6,0.067,0.013}
\definecolor{emph2}{rgb}{0.18,0.18,0.60}

\title{Adjoint SU(2) with Four Fermion Interactions}

\ShortTitle{ Four Fermion Interactions with Wilson Fermions}

\author{\speaker{Jarno Rantaharju}\\
       Department of Physics, Duke University\\
       E-mail: \email{jmr108@phy.duke.edu}}

\author{Vincent Drach\\
        CERN, Physics Department, 1211 Geneva 23, Switzerland\\
        E-mail: \email{vincent.drach@cern.ch}}
        
\author{Claudio Pica\\
       CP3 -Origins \& IMADA, University of Southern Denmark\\
       E-mail: \email{pica@cp3.sdu.dk}}
       
\author{Francesco Sannino\\
       CP3 -Origins and the Danish IAS,  University of Southern Denmark\\
       E-mail: \email{sannino@cp3.sdu.dk}}

\abstract{
Four fermion interactions appear in many models of Beyond Standard Model physics. In Technicolour and composite Higgs models Standard Model fermion masses can be generated by four fermion terms. They are also expected to modify the dynamics of the new strongly interacting sector. In particular in technicolour models it has been suggested that they can be used to break infrared conformality and produce a walking theory with a large mass anomalous dimension. We study the SU(2) gauge theory with 2 adjoint fermions and a chirally symmetric four fermion term. We demonstrate chiral symmetry breaking at large four fermion coupling and study the phase diagram of the model.
}

\FullConference{The 34th International Symposium on Lattice Field Theory\\
		 24-30 July 2016\\
		 University of Southampton, UK}

\begin{document}


\section{Introduction}

Recent studies show that apparently perturbative Higgs Yukawa models, similar to the Standard Model, can abide compositeness conditions \cite{Krog:2015bca,Bardeen:1989ds,Chivukula:1992pm,Bardeen:1993pj}. In a certain region of parameter space, the high energy description of the model does not include a propagating Higgs like state. Instead, the theory becomes a gauged NJL model \cite{Nambu:1961fr}.

Four fermion interactions are a natural part of both Technicolor \cite{Weinberg:1975gm,Susskind:1978ms} and Composite Higgs \cite{Kaplan:1983fs,Kaplan:1983sm} models. They appear as an effective description of a more complete theory of fermion mass generation. The terms connecting the Higgs sector and the top quark are usually seen as being produced by a high energy gauge or scalar interaction. A detailed example of both in which a more fundamental theory consisting of only fermions generates such terms in a model unifying both Technicolor and Composite Higgs has been described in \cite{Cacciapaglia:2015yra}.

A high energy interaction connecting the Standard Model and the Higgs sector will generate three types of four fermion terms:
\begin{align*}
&L_\text{eff} = \frac{a}{\Lambda^2_{UV} } (\bar\Psi_{SM}\Psi_{SM})^2  + \frac{b}{\Lambda^2_{UV} } \bar\Psi_{SM}\Psi_{SM}\bar\Psi_{TC} \Psi_{TC}  + \frac{c}{\Lambda^2_{UV} } (\bar\Psi_{TC}\Psi_{TC})^2.
\end{align*}
While the first term, involving only Standard Model fermions, is suppressed by the cut off scale $\Lambda_{UV}$, the other two terms may be enhanced by the dynamics of the technicolour sector. 
As was suggested in \cite{Holdom:1981rm}, the fermion mass term can be enhanced dynamically in a model with walking dynamics and a large mass anomalous dimension. This may be achieved in a natural way by having the third, NJL type term induce chiral symmetry breaking in a model that is otherwise infrared conformal \cite{Fukano:2010yv,Yamawaki:1996vr}.

In previous work we have studied the NJL model in the absence of a gauge interaction \cite{Rantaharju:2015nep,Rantaharju:2016jxy}. The results are in qualitative agreement with meanfield calculations and we are able to establish chiral symmetry breaking above a critical four fermion coupling.
In this work we focus on the SU(2) gauge theory with 2 fermions in the adjoint representation. The model has been studied extensively in the absence of a four fermion term and appears infrared conformal \cite{Catterall:2008qk,
    DeGrand:2011qd, Patella:2012da,
    Giedt:2012rj, Bergner:2015jdn,
    Rantaharju:2015yva,DelDebbio:2015byq,Rantaharju:2015cne}.
We study the phase diagram of the model at a constant gauge coupling $\beta=2.25$ and find chiral symmetry breaking above a critical four fermion coupling $g=0.25$. We study the order of the transition and, below the critical coupling, the anomalous dimension using the hyperscaling relation.

\section{The Model}

We study the SU(2) gauge theory with 2 fermions in the adjoint representation and a NJL type four fermion interaction with a partially conserved chiral symmetry. The model is defined by the action
\begin{align}
 S &= \beta_L \sum_{x,\mu<\nu} L_{x,\mu\nu}(U) + \sum_{x,y} \bar\Psi(x) D_W(x,y) \Psi(y) + \sum_{x} m_0 \bar\Psi(x) \Psi(x) \label{action} \\
  &- \sum_{x} a^2 g^2 \left [ \bar\Psi(x)\Psi(x)\bar\Psi(x)\Psi(x) + \bar\Psi(x) i\gamma_5\lambda^a\Psi(x)\bar\Psi(x) i\gamma_5 \lambda^a\Psi(x) \right ], 
\end{align}
where $L_{x,\mu\nu}(U)$ is the Wilson plaquette gauge action, $D_W$ and the Wilson Dirac operator and $a$ is the lattice spacing. In order to integrate out the fermion field, we rewrite the action using auxiliary fields:
\begin{align}
 S &=  \beta_L \sum_{x,\mu<\nu} L_{x,\mu\nu}(U) + \sum_{x} \bar\Psi(x) \left [ D_W + m_0+\sigma(x) + \pi_3(x) i \gamma_5\tau_3 \right ] \Psi(x) +\frac{\sigma(x)^2+\pi_3(x)^2}{4a^2g^2}. \label{actionaux} 
 \end{align}
The original action is recovered by integrating over the auxiliary fields.
It is now straightforward to produce configurations of $U$, $\sigma$ and $\pi_3$ using the HMC algorithm.

The four fermion interaction preserves a U(1) subgroup of the original SU(4) chiral flavour symmetry group\footnote{It is possible to build a four fermion term that preserves a SU(2)$\times$SU(2) subgroup, but in this case the auxiliary field representation does not produce a positive fermion determinant. Consequently the HMC algorithm cannot be used to generate configuration. }. As result the pseudoscalar meson spectrum is split into a single diagonal state and four degenerate non-diagonal states. When the chiral symmetry is broken, the diagonal state becomes a massless pseudo Nambu-Goldstone boson, while the non-diagonal states gain a mass from the condensate. The axial current is only conserved in the diagonal direction.

The mechanism for the breaking of the non-diagonal axial current is interesting and useful for determining the chiral condensate. We can gain insight into the chiral symmetry by writing down the PCAC relation for different flavour components:
\begin{align}
\partial_\mu \ev{ A^{I,d}_\mu(x) O } = 2\bar m \ev{ P^d(x) O } - a^2 \bar g^2 \left( \delta^{d,1} + \delta^{d,2} \right ) \ev{ S^0(x) P^d(x) O }.
\end{align}
For convenience we have already absorbed all order $1$ and order $a$ terms into a renormalized axial current \cite{Luscher:1996vw}. Thus
\begin{align}
A_\mu^{I,d}(x) &= Z_A \bar\Psi(x) \gamma_\mu \gamma_5 \tau^d \Psi(x) + c_A \partial_\mu P^d(x),\\
P^d(x) &= \bar \Psi(x) \gamma_5 \tau^d \Psi(x) \textrm{ and } S^0(x) = \bar\Psi(x) \Psi(x),
\end{align}
where $\tau^d$ are Pauli matrices in the flavour space\footnote{Only the three axial flavour transformations produce currents of this form. A similar argument holds for the remaining 2 directions.}.
In this study we neglect any correction arising at order $a$ or higher, including $c_A$.
Chiral symmetry is naturally restored when $\bar m=0$ and thus $\partial_\mu\ev{A^{I,3}_\mu }=0$.

The variation of the four fermion term is nonzero when $d=1$ or $2$. This is naturally the source of the symmetry breaking, but it is an order $a^2$ term. If chiral symmetry is broken and the dimensionless quantity $a^3 \bar\Psi(x)\Psi(x)$ has a nonzero expectation value
\begin{align}
\Sigma_L = \frac{a^3}V \sum_x \ev{ \bar\Psi(x)\Psi(x)} \neq 0,
\end{align}
we can rewrite the term as using a subtracted scalar density $S_S^0(x) = S^0(x) - \Sigma_L/a^3$:
\begin{align}
&\frac{1}{a} \bar g^2 \Sigma_L \left( \delta^{d,1} + \delta^{d,2} \right ) \ev{P^d(x) O } + a^2 \bar g^2 \left( \delta^{d,1} + \delta^{d,2} \right ) \ev{ S_S^0(x) P^d(x) O }.
\end{align}
The additional mass like term explains how the axial current can remain broken in the non-diagonal directions even when $\bar m=0$. The combination $\bar g^2 \Sigma_L$ does not receive additive renormalization and is only nonzero if the chiral symmetry is broken. It is useful as an order parameter for chiral symmetry breaking.

\section{The Phase Diagram}

We study the phase diagram with a constant gauge coupling $\beta=2.25$. As in the ungauged model, we expect the chiral symmetry to be broken at a large enough value coupling $g$. In the ladder approximation \cite{Yamawaki:1996vr} the critical coupling is 
\begin{align*}
g_c &= \frac{1}{2}\left(1+\sqrt{1-\frac{\lambda}{\lambda_c}} \right )g^*,
\end{align*}
where $g^*$ is the critical coupling at zero gauge coupling. We can therefore expect a smaller critical coupling than the pure NJL results, $g\approx 0.45$.

\begin{figure}
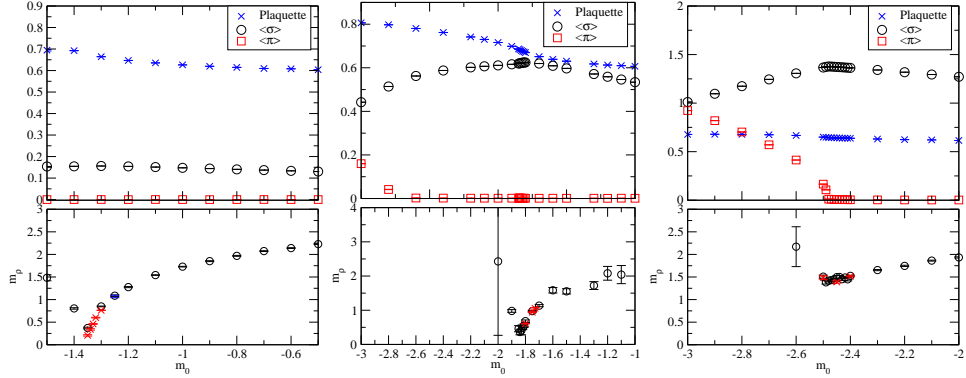
 \center
\includegraphics[height=0.33\linewidth]{{vmass_auxave_s0.1_b2.25}.eps}
\includegraphics[height=0.33\linewidth]{{vmass_auxave_s0.2_b2.25}.eps}
\includegraphics[height=0.33\linewidth]{{vmass_auxave_s0.3_b2.25}.eps}
\caption{Scans of the phase space at $g=0.1,0.2$ and $0.3$ from left to right respectively. In the lower plot depicting the vector meson mass, the red points are measured with the lattice size $V=24^3\times64$ and the blue points with $V=18^3\times 36$. All other measurements done are with $V=16^4$. At $g=0.1$ and $0.2$ we find a critical point where the vector meson mass $m_\rho$ approaches zero. There is no visible change in $\ev{\pi}$ at the critical line, but the behaviour of $\ev{\sigma}$ and the plaquette do change. At $g=0.3$ the vector meson mass never reaches zero. The pseudoscalar density $\ev{\pi}$ acquires an expectation value, indicating a critical point. The nonzero vector meson mass indicates chiral symmetry breaking. }
\label{phase scans}
\end{figure}

In order to find the expected phase transitions we measure the expectation values of the auxiliary fields,
\begin{align*}
 &\ev{\sigma} = \frac{1}{V} \ev{\sum_x  \sigma(x) } = - \frac{2a^2g^2}{V} \,\ev{\sum_x \bar\Psi(x)\Psi(x)} \\
 &\ev{\pi} = \frac{1}{V} \ev{\sum_x  \pi_3(x) }  = - \frac{2a^2g^2}{V} \ev{\sum_x\bar\Psi(x)  i\gamma_5 \tau_3 \Psi(x)},
\end{align*}
the plaquette expectation value and the mass of the diagonal vector meson $m_\rho$. The results are shown in figure \ref{phase scans} with three four fermion couplings, $g=0.1,0.2$ and $0.3$. 
They are qualitatively similar to the pure NJL case \cite{Rantaharju:2016jxy}. In the unbroken, infrared conformal phase we expect to find a primary critical line where all masses approach zero. At $g=0.1$ and $0.2$ we find a critical line where $m_\rho$ approaches zero and the behaviour of the other measurables change. At $g=0.2$ we also see a secondary critical line on the negative mass side, where $\ev\pi$ becomes nonzero. At $g=0.3$, $\ev\pi$ becomes nonzero at the primary critical line and $m_\rho$ remains nonzero. The transition appears to be second order implying a divergence in the correlation length of the diagonal pseudoscalar density and therefore zero mass for the diagonal pseudoscalar meson. The nonzero vector meson mass indicates chiral symmetry breaking.

\begin{figure}
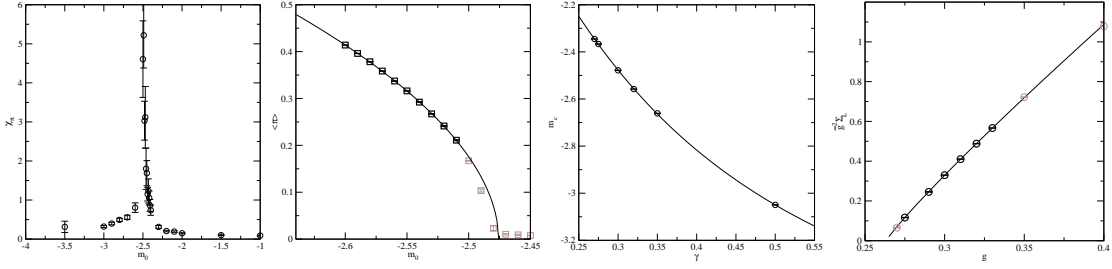
 \center
\includegraphics[height=0.23\linewidth]{{pi_susc_s0.3_b2.25}.eps}
\includegraphics[height=0.23\linewidth]{{pi_ave_fit_s0.3_b2.25}.eps}
\includegraphics[height=0.23\linewidth]{{mcrit}.eps}
\includegraphics[height=0.23\linewidth]{{gsigma_fit}.eps}
\caption{ From left to right: Scaling fits to the susceptibility $\chi_\pi=\ev{\pi^2}-\ev{\pi}^2$ and the expectation value $\ev{\pi}$ with $\beta=2.25$ and $\gamma=0.3a$. The critical $m_c(g)$ line in the chirally broken phase and the order parameter $ \bar g^2 \Sigma_L$ in the chirally broken phase. }
\label{gauged_pi_susc}
\end{figure}

To verify the latter point we measure the susceptibility of $\chi_\pi = \ev{\pi^2}-\ev{\pi}^2$, shown in the first panel of figure \ref{gauged_pi_susc}. There is indeed a peak in the susceptibility at the critical line. In addition, we show a scaling fit to
\begin{align}
\ev\pi = C \left( m_c - m_0 \right)^\beta \label{scalingfit}
\end{align}
in the second panel of figure \ref{gauged_pi_susc} and find $\beta=0.525(8)$ with a $\chi^2/d.o.f. = 0.67$. This is compatible with a second order transition in the meanfield universality class.

We then proceed to determine the order of the chiral symmetry breaking transition by studying the order parameter $ \bar g^2 \Sigma_L$ along the critical line. This requires finding the critical point $m_c(g)$ with multiple values of $g$ by fitting to the scaling relation \ref{scalingfit}. 
The third panel of \ref{gauged_pi_susc} shows the resulting values of $m_c(g)$ with a second order fit in $1/g$. We then measure $ \bar g^2 \Sigma_L$ at several values of the coupling along the critical line and fit to the scaling relation
\begin{align}
\bar g^2 \Sigma_L = C_g \left( g - g_c \right)^{\beta_g}. \label{Sigmascalingfit}
\end{align}
Here we find a $\chi^2/d.o.f$ of 1.16 and $\beta_g = 0.909(7)$, $g_c = 0.2633(2)$ and $C_g=3.536(4)$. The fit is shown in the fourth panel of figure \ref{gauged_pi_susc}.

\section{The Anomalous Dimension}

One of the main motivations for studying the model is that the four fermion interaction is expected to increase the mass anomalous dimension along the infrared fixed line of the chirally symmetric phase. In order to verify the expectation and to quantify the effect we measure the anomalous dimension at $g=0.1$ and $g=0.2$ using the hyperscaling relation.

\begin{figure}
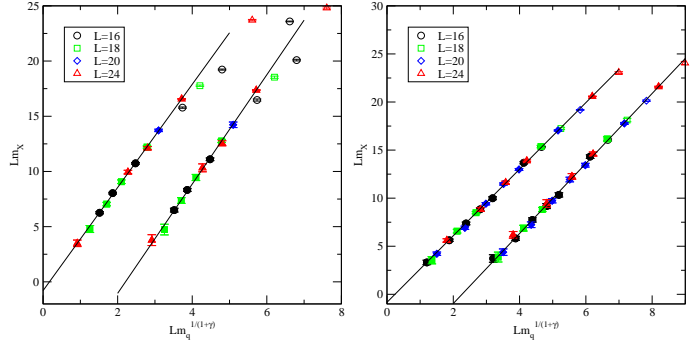
 \center
\includegraphics[height=0.3\linewidth]{{nd_combined_scaling_fit_s0.1}.eps}
\includegraphics[height=0.3\linewidth]{{nd_combined_scaling_fit_s0.2}.eps}
\caption{The conformal hyperscaling fit to the non-diagonal pseudoscalar and vector meson masses at $g=0.1$ and $g=0.2$ respectively. The points with filled symbols are included in the fit. The vector mass is shifted to the left}
\label{hyperscaling}
\end{figure}

At a nonzero mass in the vicinity of a conformal critical point all masses scale with the quark mass with the same scaling exponent,
\begin{align*}
  &Lm_X = f(x) = a_X x + c_X \\
  &x= \left | m_0-m_c \right |^\frac{1}{1+\gamma_m}.
\end{align*}
We measure the diagonal pseudoscalar and vector masses with several lattice sizes and bare masses. The fits are shown in figure \ref{hyperscaling}. At $g=0.1$ we find the best fit is $m_c=-1.357(1)$ and $\gamma_m = 0.54(6)$ with a $\chi^2/d.o.f. = 0.6$. However the fit is not robust and by varying the fit range we find $0.4 < \gamma_m < 0.6$. At $g=0.2$ the best fit is $m_c = -1.8276(5)$ and $\gamma_m = 0.89(3)$ with a $\chi^2/d.o.f. = 0.9$. Varying the fit range yields $0.6 < \gamma_m < 0.9$. Both results are larger than the values measured in the absence of a four fermion term, $\gamma_m \sim 0.3 - 0.4$ \cite{DeGrand:2011qd,Patella:2012da,Rantaharju:2015yva,Rantaharju:2015cne}.

\section{Conclusions}

We report a preliminary study of the SU(2) gauge model with 2 fermions in the adjoint representation and a NJL type four fermion interaction. The model is an approximate realization of a walking technicolour model in which infrared conformality is broken by a four fermion interaction. Such models are expected produce a large mass anomalous dimension together with a slowly running coupling.

We study the phase diagram and the mass anomalous dimension in the infrared conformal phase. The anomalous dimension is measured with a single gauge coupling $\beta=2.25$ and furthermore the phase diagram is studied only at the lattice size $L=16$. Nevertheless our results confirm the qualitative predictions obtained using the ladder approximation \cite{Yamawaki:1996vr} and mean field theory. 

Chiral symmetry is spontaneously broken above a critical four fermion coupling. We are able to measure the order parameter of chiral symmetry breaking $ \bar g^2\Sigma_L$ accurately and find a result consistent with a second order transition.
We measure the mass anomalous dimension at two values of the coupling in the infrared conformal phase. The systematic errors are large, but the measurement seems consistent with expectations. The anomalous dimension increases with $g$ along the critical line.

\section{Acknowledgements}

This work was supported by the Danish National Research Foundation DNRF:90 grant and by a Lundbeck Foundation Fellowship grant. The computing facilities were provided by the Danish Centre for Scientific Computing and the DeIC national HPC center at SDU.

\end{document}